\documentclass{article}
\usepackage{dcase2023,amsmath,graphicx,url,times,booktabs, tabularx}

\title{TEXT-DRIVEN FOLEY SOUND GENERATION WITH LATENT DIFFUSION MODEL}

\name{Yi Yuan,
      Haohe Liu,
      Xubo Liu, 
      Xiyuan Kang,
      Peipei Wu,
      Mark D. Plumbley, 
      Wenwu Wang
      }

\address{Centre for Vision Speech and Signal Processing, University of Surrey, United Kingdom\\          
 }
\begin{document}

\ninept
\maketitle

\begin{sloppy}

\begin{abstract}
Foley sound generation aims to synthesise the background sound for multimedia content. Previous models usually employ a large development set with labels as input~(e.g., single numbers or one-hot vector). In this work, we propose a diffusion model based system for Foley sound generation with text conditions. To alleviate the data scarcity issue, our model is initially pre-trained with large-scale datasets and fine-tuned to this task via transfer learning using the contrastive language-audio pretraining~(CLAP) technique. We have observed that the feature embedding extracted by the text encoder can significantly affect the performance of the generation model. Hence, we introduce a trainable layer after the encoder to improve the text embedding produced by the encoder. In addition, we further refine the generated waveform by generating multiple candidate audio clips simultaneously and selecting the best one, which is determined in terms of the similarity score between the embedding of the candidate clips and the embedding of the target text label. Using the proposed method, our system ranks ${1}^{st}$ among the systems submitted to DCASE Challenge 2023 Task 7. The results of the ablation studies illustrate that the proposed techniques significantly improve sound generation performance. The codes for implementing the proposed system are available at~\url {https://github.com/yyua8222/Dcase2023_task7}.  
\end{abstract}


\section{Introduction}
\label{sec:intro}

The development of deep learning models has recently achieved remarkable breakthroughs in the field of sound generation~\cite{audioldm,audiolm,makeaudio,audiogen}. Among various application domains of sound, Foley sounds, the mimic of background sound, play a crucial role in enhancing the perceived acoustic properties of movies, music, videos and other multimedia content~\cite{dcaset7}. The development of an automatic Foley sound synthesis system holds immense potential in simplifying traditional sound generation, which often involves intensive labour work on sound recording and mixing.

Currently, many sound generation models~\cite{audioldm,diffsound,Liu-tts} adopt an encoder-decoder architecture, showing remarkable performance. Liu et al.~\cite{Liu-tts} utilize a convolutional neural network~(CNN) encoder, a variational autoencoder~(VAE) decoder and a generative adversarial network~(GAN) vocoder. The encoder embeds the input feature~(e.g., label) into latent variables and the decoder transforms this intermediate information into mel-spectrogram which is then converted to a waveform by the vocoder. Diffsound~\cite{diffsound} utilizes text as input and obtains the semantic features by using a contrastive language image pre-training~(CLIP) model~\cite{clip}. AudioGen~\cite{audiogen} further improves the performance by using a pre-trained Transfer Text-to-Text Transformer~(T5)~\cite{t5} to obtain text embedding, which is then used to generate the waveform directly without using a vocoder. 

\begin{figure}[ht]
    \centering
    \includegraphics[width=0.83\linewidth]{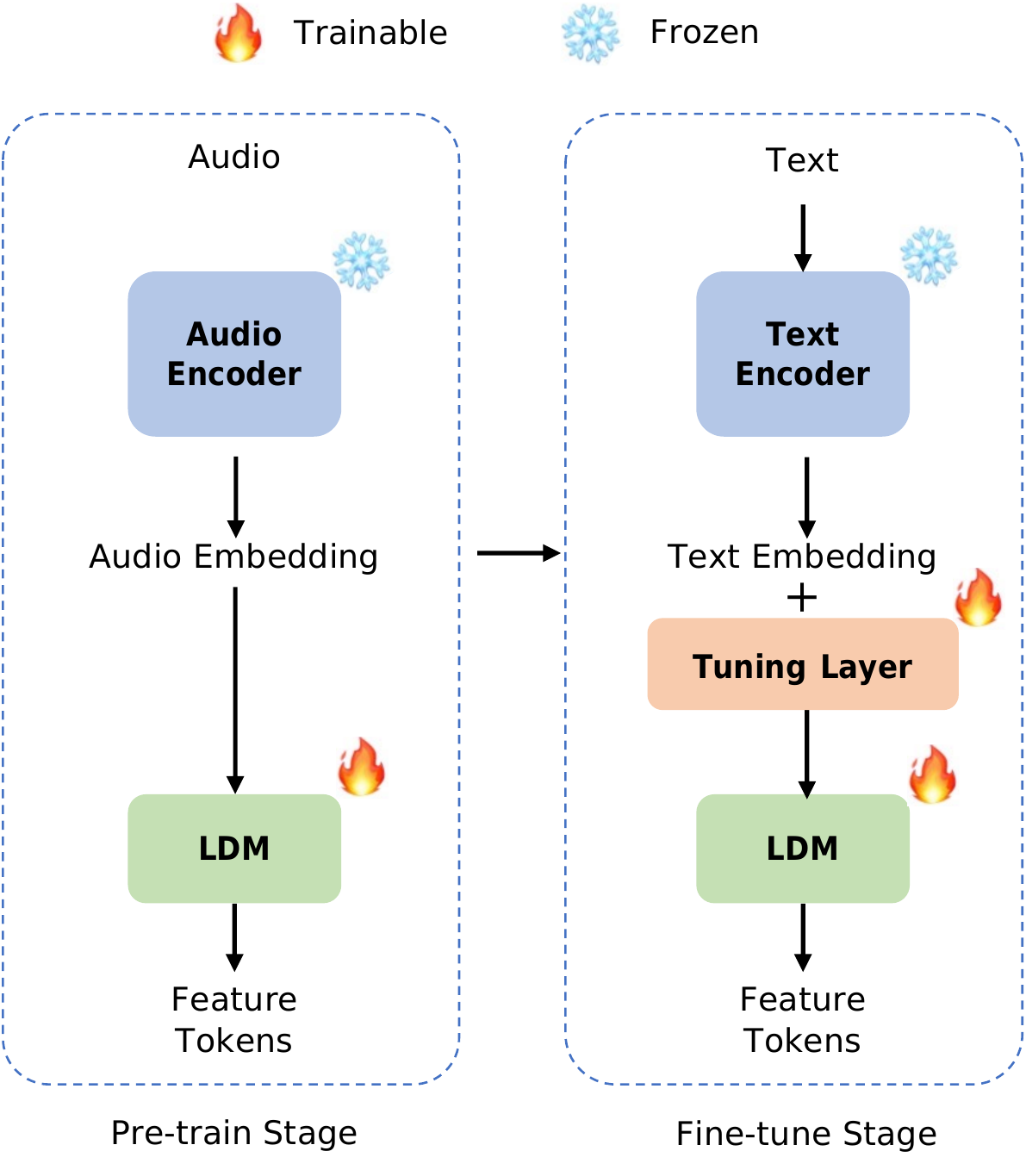}
    \caption{The training process of the LDM model}
    \label{fig:tuning}
\vspace{-2mm}
\end{figure}

\begin{figure*}[htbp]
    \centering
    \includegraphics[width=\linewidth]{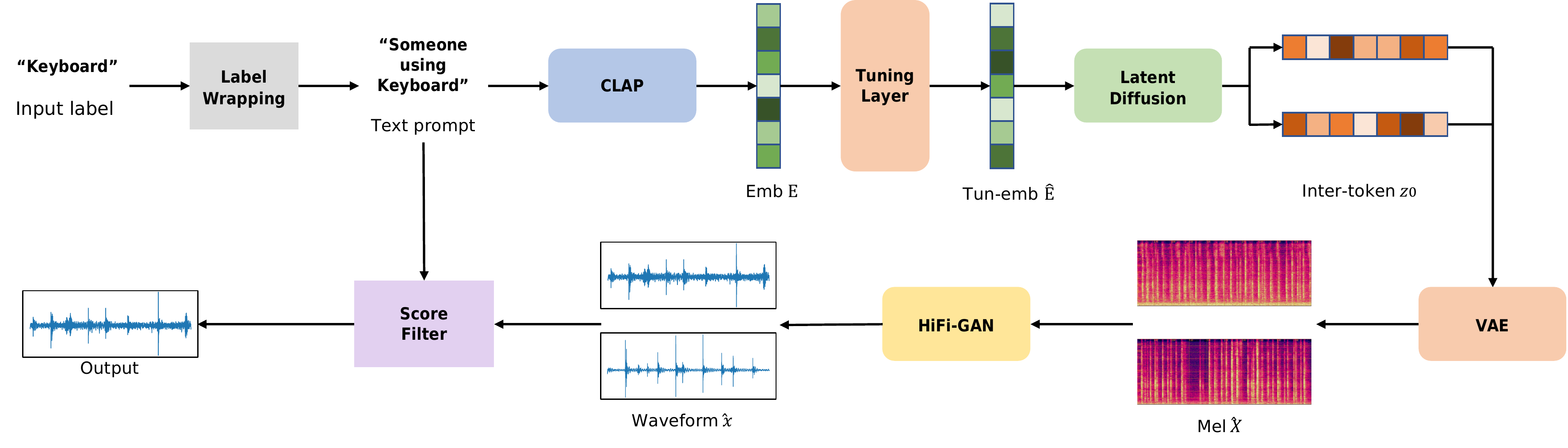}
    \caption{The overview of the system}
    \label{fig:overview}
\end{figure*}

This paper proposes a latent diffusion model~(LDM) based method for Foley sound generation. Our model follows the structure of AudioLDM~\cite{audioldm}, an audio generation model that comprises a diffusion model based encoder, a VAE based module for learning audio prior, and a HiFi-GAN vocoder for waveform generation. Due to the lack of training data for the sound generation task, we follow the idea of pre-training~\cite{yuan2023,pre-training1}, by initially training all three models on large-scale datasets such as AudioSet~\cite{audioset}, AudioCaps~\cite{audiocaps} and Freesound\footnote{https://freesound.org}, and then transferring them onto the target development set. For inputs, the category labels are initially wrapped into relevant texts~(e.g., turning the label “Keyboard” into text “Someone using keyboard”) before they are passed into the contrastive language-audio pre-training~(CLAP)~\cite{clap} for generating the text embeddings. To learn the most suitable semantic features of each sound, an embedding tuning layer is then added to text embedding for finding the optimal embedding during the fine-tuning stage. As shown in Figure~\ref{fig:tuning}, we first use audio embeddings to pre-train the LDM model. Then a tuning layer is introduced into the system, which is updated via transfer learning along with the LDM module. For outputs, the cosine-similarity score obtained in terms of the outputs and target labels is used to select the best-related sounds from a pool of candidate sound clips, which can improve the overall quality of the final result. Through experiments and ablation studies, we observe that the proposed techniques in our system can significantly improve the model performance in both the relevance of generated sound and the stability of the overall quality. Our system achieves a Fréchet audio distance~(FAD) score of $4.52$ on the DCASE task 7 validation set, significantly better than the baseline model with a FAD of $9.7$.

The remaining sections of this paper are organised as follows. Section 2 describes the overview of the proposed system. The methodology of the network is explained in Section 3. Section 4 introduces the experimental setup. Results are shown in Section 5. Section 6 summarizes this work and draws the conclusion.

\section{SYSTEM OVERVIEW}

\label{sec:format}

Our proposed system is based on the widely used structure on sound generation, which consists of an encoder, a generator, a decoder and a vocoder. The system adopts the same structure as AudioLDM~\cite{audioldm}, which used the CLAP~\cite{clap} as the encoder and a latent diffusion model as the generator. 

As a cascade model, the decoder and vocoder are trained separately and then built into the overall system with the trained parameters frozen when training the LDM model based generator. Instead of directly using labels as the input, we employ a wrapping strategy to generate text descriptions for each label as the initial mechanism to enhance the semantic information of the input. For example, we turn the label “Keyboard” into texts “Someone using keyboard”. Then, we introduce an embedding tuning layer after the encoder in order to produce a more suitable embedding for each sound. 

During the generating stage, with the text input, the system extracts the text embedding using the CLAP model, and the LDM model then generates the intermediate representation of the sound feature, using the text embedding as a condition. Subsequently, the mel-spectrogram can be decoded from the tokens by the VAE decoder, which is then transformed into waveform by the GAN vocoder. This system is then further improved with several techniques:
\begin{itemize}
    \item Transfer learning is introduced to boost the performance by pre-training the model on larger datasets.
    \item A tuning layer is applied during the fine-tuning stage to find the optimal embedding.
    \item Similarity score between the embedding of the generated output and the target embedding is applied to select the best match results among a group of waveform clips generated by the system. 
\end{itemize}
Detailed explanations of these methods are provided in the following section. The overall sampling procedure is shown in Fig. \ref{fig:overview}.

\section{PROPOSED METHOD}

\subsection{System structure}
\label{clap}
\subsubsection{CLAP based encoder} 
We use the Contrastive Language-Audio Pretraining~(CLAP)~\cite{clap} model to obtain the embedding of the input. CLAP consists of a text encoder $\textit{f}_{text}$ that turns a text description \textit{y} into text embedding $\boldsymbol{E}^{y}$ and an audio encoder $\textit{f}_{audio}$ that computes an audio embedding $\boldsymbol{E}^{x}$ from audio samples \textit{x}. The two encoders are trained with cross-entropy loss, resulting in an aligned latent space with the same dimension $\textit{D}_{e}$ for both audio and text embedding. Since most large audio datasets (e.g., AudioSet) only provide audio-label pairs, we leverage the cross-modal information provided by two encoders. Specifically, the system is pre-trained on larger datasets with audio embedding and fine-tuned with text embedding on the task development set. During the fine-tuning process, the text embedding $\boldsymbol{E}^{y}$ is passed through a trainable linear layer to find the optimal embedding feature for each class of sound. Details of this mechanism are presented in Section~\ref{sec:tuning}

\subsubsection{LDM based generator} Our system uses an LDM~\cite{stable_diffusion} to generate the intermediate latent tokens, with the feature embedding ($\boldsymbol{E}^{y}$ or $\boldsymbol{E}^{x}$) as the condition. These tokens are then used by the VAE decoder to generate the mel-spectrogram. During training, the LDM involves two processes: 1) A forward process where the latent vector $\boldsymbol{z}_{0}$ is gradually turned into a standard Gaussian distribution $\boldsymbol{z}_{N}$ in $\textit{N}$ steps, with noise $\boldsymbol{\epsilon}$ added in each step. 2) A reverse process for the model to predict the transition probabilities $\boldsymbol{\epsilon}_{\theta}$ of each step \textit{n}, for reconstructing the data $\boldsymbol{z}_{0}$ by removing the noise $\boldsymbol{z}_{N}$. 
The model is trained with a re-weighted objective~\cite{DDPM} as:

\begin{equation}
    L_{n}(\theta)={E}_{\boldsymbol{z}_{0},\boldsymbol{\epsilon},n}|| \boldsymbol{\epsilon} - \boldsymbol{\epsilon}_{\theta}(\boldsymbol{z}_{n},n,\boldsymbol{E})||^2_{2}
\label{ldm_loss}
\end{equation}
where $\boldsymbol{\epsilon}_{\theta}$ is the Gaussian distribution predicted by LDM with current state $\boldsymbol{z}_{n}$, current step $n$, and current condition $\boldsymbol{E}$. 
During sampling, the model first generates random Gaussian noise as $\textit{z}_{N}$, and then applies the denoising process by predicting the reverse transition probability and taking the $\boldsymbol{E}$ from CLAP as the condition. 

\subsubsection{VAE decoder \& HiFi-GAN vocoder} We utilize a combination of a VAE decoder and a HiFi-GAN vocoder to transform latent feature tokens into waveforms. Our approach involves training a VAE~\cite{specvqgan} to decode the latent feature tokens into mel-spectrograms, and a HiFi-GAN~\cite{hifigan} to generate the corresponding waveforms. To achieve this, we initially convert the waveforms into mel-spectrograms using the Short-Time Fourier Transformation (STFT). The VAE is trained to compress the mel-spectrograms, $\boldsymbol{X}$, into a latent space vector $\textit{z}_{0}$, then reconstruct the mel-spectrograms $\hat{\boldsymbol{X}}$ from the compressed representation. In parallel, we employ a HiFi-GAN to convert the mel-spectrograms $\hat{\boldsymbol{X}}$ into the corresponding waveform representations, denoted as $\hat{\textit{x}}$. 

\subsection{Practical issues}
\noindent \textbf{Transfer learning}
To deal with the issue of data scarcity, our system takes advantage of a pre-trained model~\cite{yuan2023} by initially training all three models on extensive audio datasets, followed by fine-tuning them on our development dataset. Specifically, the LDM model undergoes its initial training phase using large-scale datasets with audio embeddings as inputs, while the model is then trained on the development dataset utilizing text embeddings. 

\noindent \textbf{Embedding tuning}
\label{sec:tuning}
To first initialize the text embedding with more semantic features, we apply some hand-picked text by extending the label with some adjunct word~(e.g., dogbark into a dog bark). We then apply a tuning strategy to determine the optimal embedding of each sound class. To implement this, we introduce a linear layer $\textit{L(x)}$ with a trainable parameter to fine-tune the text embedding before passing it to the LDM model. To guide this trainable layer with only minor updates on the embedding, the parameters are initialized with an identity matrix as weight, along with a Gaussian noise as bias $\textit{l}_{b}$. Hence, the initial $\textit{L(x)}$ serves as an adding function of input $\textit{x}$ and bias $\textit{l}_{b}$ at the beginning of the training process. Then, the system learns to update the parameter of both weight and bias for optimal embedding during training. The embedding updated by this linear layer is also used as the target embedding for the score-selecting function discussed in the following section. 

\noindent \textbf{Score-based selection}
To improve the overall generation quality and robustness, a scoring mechanism is applied to determine the best matches among sampling results. Leveraging the fact that CLAP provides embeddings in the same latent space for audio and text, we utilize the cosine similarity between the output audio and the target text. By comparing the FAD score of different groups of output clips with different score-selecting thresholds, specific thresholds are established for each class, allowing the system only selects the results surpassing these thresholds. 


\begin{table*}[htbp]
\centering
\begin{tabular}[\linewidth]{cccccccc}
\hline
\multicolumn{1}{c}{System}        & Dog Bark      & Footstep     & Gun Shot       & Keyboard     & Moving Motor Vehicle & Rain          & Sneeze Cough  \\ \hline
\multicolumn{1}{c}{Basline~\cite{Liu-tts}}       & $13.41$        & $8.11$         & $7.95$          & $5.23$        & $16.11$               & $13.34$        & $3.77$          \\
\hline
\multicolumn{1}{c}{LDM-S} & $4.41$          & $7.44$          & $7.46$           & $3.13$         & $16.97$                & $12.62$         & $3.02$          \\
\multicolumn{1}{c}{LDM-S+Pre} & $4.17$          & $6.86$          & $7.25$           & $3.15$         & $15.68$                & $12.95$         & $2.85$          \\
\multicolumn{1}{c}{LDM-S+Pre+Text}  & $3.84$          & $5.66$          & $6.66$           & $3.48$         & $14.35$                & $12.62$         & $2.12$          \\ 
LDM-S+Pre+Text+Filter                     & $3.53$ & $5.04$ & $5.65$ & $2.80$ & $15.29$                & $9.76$          & $\textbf{1.92}$ \\
LDM-S+Pre+Text+Filter+Tuned                     & $\textbf{3.36}$ & $\textbf{4.77}$ & $\textbf{5.19}$ & $\textbf{2.69}$ & $14.83$                & $10.00$          & $1.98$ \\
LDM-L+Pre+Text+Filter+Tuned                   & $6.04$          & $5.05$          & $6.44$           & $3.07$         & $\textbf{11.08}$       & $\textbf{4.74}$ & $2.93$  \\
\hline
\end{tabular}
\caption{The best results of each system on the DCASE2023-T7 evaluation set. LDM-S: model trained from scratch. Pre: model with pre-training on large datasets. Text: using label-related text as input. Filter: applying the score-selecting function. Tuned: model with a fine-tuned embedding. The score selection for motor sound is used with the text embedding of “ A moving motor ”.}
\label{table:results}
\vspace{-2mm}
\end{table*}

\section{EXPERIMENTAL SETUP}
\subsection{Dataset}
\textbf{DCASE2023-T7} consists of a training set and an official evaluation set with seven different classes of fully labelled urban sounds.  Each class has around $600$ to $800$ $4$-second sound clips in the training set and exactly $100$ clips in the evaluation set. We randomly partitioned the training dataset into two subsets, with a ratio of $9:1$ for training and validation purposes, while the evaluation set was exclusively used during the evaluation phase.

\noindent \textbf{AudioSet} is a large-scale dataset for audio research, which consists a wide range of sounds. In detail, Audioset provides around $2.1$ million 10-second audio with $527$ classes of labels. Our system uses AudioSet during the pre-training stage. 

\noindent \textbf{Freesound} is a similar audio dataset with labels but with a non-fixed length, ranging from one second to several minutes. To unify the output length, all the sounds in Freesound are padded into a 10-second-long clip to match the data in Audioset. 

By combining AudioSet and Freesound, we collected around $2.2$M sounds in $22.05$Khz for pre-training the LDM, VAE and GAN models. By using the audio-embedding and mel-spectrogram as input conditions, we only utilize the audio features to pre-train the models, while label features are then used during fine-tuning stage with the official training dataset. 

\subsection{Evaluation metrics}
We apply the FAD~\cite{fad} score as main evaluation metric. In detail, FAD calculates the Fréchet distance \textit{F} between a  group of target sound audio clip \textit{t} and a group of generated sound audio clip \textit{r}, formed:

\begin{equation}
\textit{F} = ||\mu_{r} - \mu_{t} ||^{2} + tr (\Sigma_{r}+\Sigma_{t} - 2\sqrt{\Sigma_{r}\Sigma_{t}})
\end{equation}
where $\mu$ and $\Sigma$ are the multivariate Gaussian of the embedding vector from each group extracted by VGGish~\cite{vggish}.  

\subsection{Parameter setting}

Both the decoder and vocoder are trained separately, then they are integrated into the overall system with parameters fixed when training the LDM model. Initially, all three models are pre-trained using AudioSet and Freesound from scratch and then fine-tuned with the development set. 

For the mel-spectrogram of $22.05$kHz sounds, we set the window length as $1024$ samples, the hop size as $256$ and the number of mel-filterbank as $80$. The VAE is trained with a compression level of $4$, which encodes the mel-spectrogram into a latent vector of $20$ in the frequency dimension and $86$ in the time dimension. The length of the audio embedding $\boldsymbol{E}^{x}$ and text embedding $\boldsymbol{E}^{y}$ from the CLAP encoder in Section~\ref{clap} is $512$. All the models are optimized with Adam optimizer under an initial learning rate of $3.0\times10^{-5}$, with $3$ epochs on the pre-training dataset and up to $1000$ epochs on the training set. We test the model~(LDM-S) performance by generating $100$ clips per class and calculate the FAD score as the evaluation metric for every $100,000$ step. 

To investigate the influence of the input embedding features, we employ a diverse set of labels and texts as embeddings for training the model. In the case of embedding tuning, we begin by selecting a specific set of text for providing the initial text embedding. Then, we train the tuning layer along with the LDM model, guiding the embedding towards the optimal value. 

To further investigate the potential of the model on sound generation, we trained a larger LDM with a bigger CLAP model~(LDM-L) with the same training configurations. To balance the computing complexity and the output quality, we trained this model on $16$kHz sounds and upsample it to $22.05$kHz before output the results. For the $16$kHz mel-spectrogram, the hop size is decreased to $160$ with a mel-bin dimension of $64$. The results of both models are shown in the following section.

\section{RESULTS AND ANALYSIS}

The performance of our system on DCASE2023-T7 validation set is reported in Table~\ref{table:results}. Most of our models outperform the baseline~\cite{Liu-tts} by a large margin in terms of FAD. The results obtained from different sizes of LDM highlight distinct strengths: LDM-S is better at generating clear and distinct sounds like dog barks, footsteps, and gunshots, whereas the larger model~(LDM-L) demonstrates superior performance in handling complex sounds such as motor sounds and rain sounds.

\begin{figure}[htbp]
    \centering
    \includegraphics[width=\linewidth]{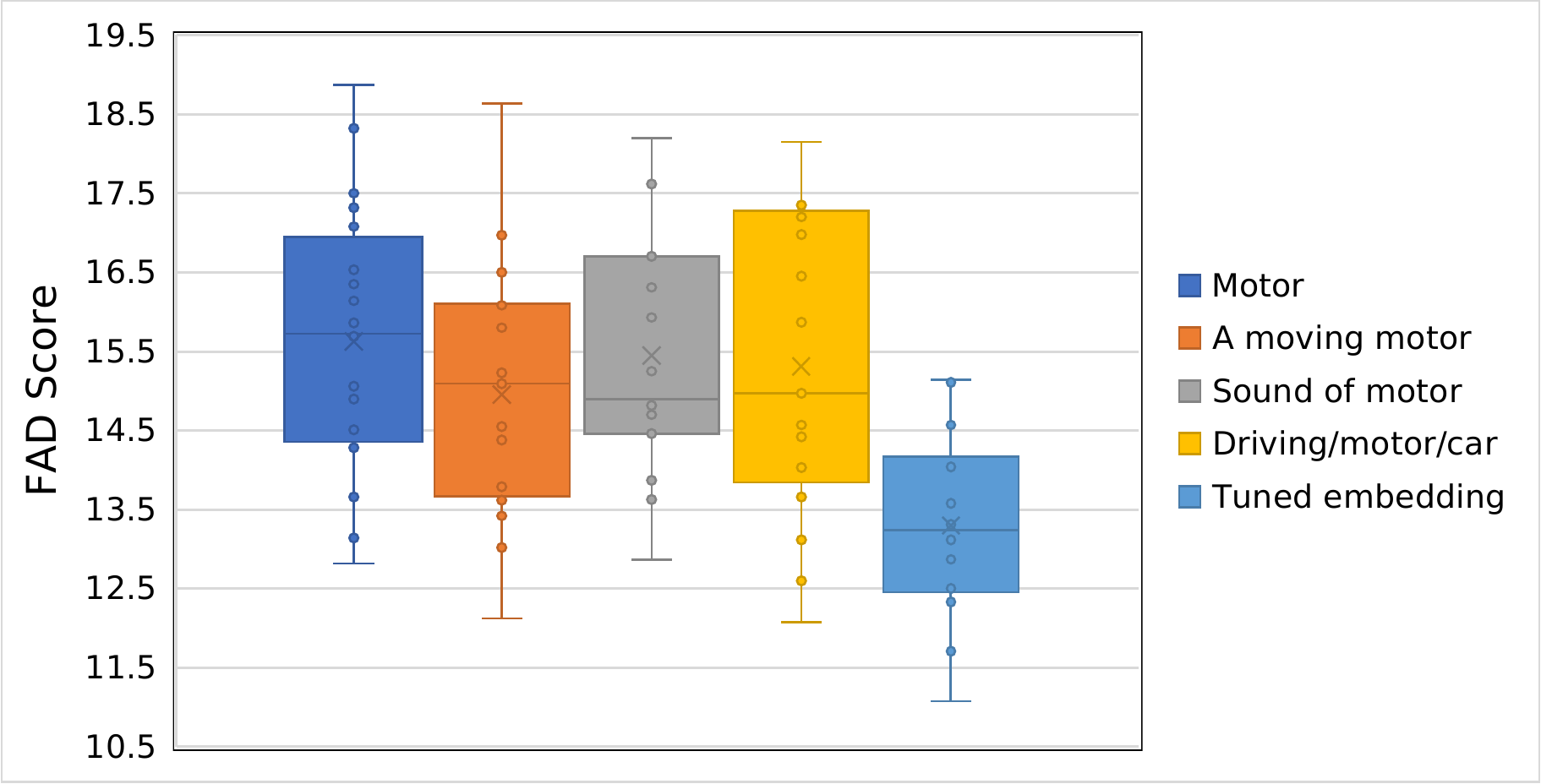}
    \caption{Results of motor with different training embedding}
    \label{fig:boxplot}
\end{figure}

Ablation studies are also conducted to investigate the effects of each proposed technique. The experimental results in Table~\ref{table:results} demonstrate that transfer learning generally improves the system performance in most cases with respect to the evaluation metrics. Applying the embedding tuning strategy enables the system to optimize the embedding value for each class, which further improves the performance. To validate the effectiveness of this embedding tuning mechanism, we conduct several experiments with models trained with different frozen text embeddings. All the models are sampled under the same configuration for up to $20$ times and the results of motor sound are presented in the box chart of Fig.~\ref{fig:boxplot}. It can be observed that training without embedding tuning may yield results with varying quality, ranging from the best FAD of around $12.5$ to the largest score of up to $18$. On the contrary, generating a well-trained embedding value can contribute to more stable results. This might be because the updated embedding during training can provide more semantic information for both LDM denoising and the waveform tuning process.

\begin{table}[ht]
\centering
\begin{tabular}{cc}
\hline
Embedding         & Moving Motor Vehicle \\
\hline
Label             & $16.97$                \\
Motor             & $13.14$                \\
A moving motor      & $12.12$                \\
Sound of motor    & $12.87$                \\
Driving/motor/car & $12.07$                \\
Tuned embedding   &  $11.08$           \\
Audio embedding  & $\textbf{8.88}$     \\
\hline
\end{tabular}
\caption{The best results on LDM-L with FAD on motor sounds between different score-selection. Embeddings indicate the text/label value for training and similarity calculation.}
  \label{table:motor}
\vspace{-2mm}
\end{table}

From Table~\ref{table:results}, the utilization of the similarity score function significantly enhances the overall performance, leading to improved output quality in most scenarios. However, despite the improvements observed in the majority of classes, we noticed that the generation quality of motor sounds did not exhibit a significant decrease in FAD~(best achieve $11.08$). By operating several subjective evaluations~(human evaluation), we find out that this might be because most motor sounds consist of noise-like sounds and sound events with distinct differences~(e.g., driving sounds and engine sounds), making it challenging for CLAP to identify and extract a single embedding that aligns perfectly with all the target clips. To address this issue and improve the correlation of the score function, we introduced a multi-target-selection approach to replace the single embedding score-selection. Specifically, we collected a set of audio embeddings that demonstrated top feature correlation with the training dataset and randomly selected an audio embedding for the score-selection during each iteration. As the result presents in Table~\ref{table:motor}, our system with multiple audio-embedding filters achieves a notable FAD score of $8.88$ for motor sounds.

\section{CONCLUSION}
This paper proposes a framework for small-domain Foley sound generation. Our system leverages a diffusion-based model and applied several methods to enhance performance. On the input feature, our experiment shows that the input embedding can significantly affect the overall quality. To alleviate this distinct gap between label and sound alignment, we proposed a trainable embedding for tuning the embedding value. Our result indicates that an improved embedding can further improve the quality and stability of the model. For output, a score-selection strategy is utilized to select the best clip along with CLAP score similarity. The experimental result shows that our system can significantly improve over the baseline network by a large margin. In the future, we will explore more efficient and end-to-end methods for audio feature extraction and fine-tuning. 
\section{ACKNOWLEDGMENT}

This research was partly supported by a research scholarship from the China Scholarship Council~(CSC) No.$202208060240$, funded from British Broadcasting Corporation Research and Development~(BBC R\&D), Engineering and Physical Sciences Research Council~(EPSRC) Grant EP/T019751/1 ``AI for Sound'', and a PhD scholarship from the Centre for Vision, Speech and Signal Processing~(CVSSP), University of Surrey. 
For the purpose of open access, the authors have applied a Creative Commons Attribution~(CC BY) license to any Author Accepted Manuscript version arising.

\bibliographystyle{IEEEtran}

\end{sloppy}
\end{document}